\documentclass[prd,aps,preprint,tightenlines,superscriptaddress]{revtex4}
\usepackage{epsfig}
\usepackage{axodraw}
\preprint{DOE/ER/40762-299} \preprint{UM-PP\#04-008}

\begin{document}
\title{Sum Rules and Spin-Dependent Polarizabilities \\
of the Deuteron in Effective Field Theory}
\author{Xiangdong Ji} \email{xji@physics.umd.edu} \affiliation{Department of
Physics, University of Maryland, College Park, Maryland 20742}
\author{Yingchuan Li}
\email{yli@physics.umd.edu} \affiliation{Department of Physics,
University of Maryland, College Park, Maryland 20742}
\date{\today}
\vspace{0.5in}
\begin{abstract}

We construct sum rules for the forward vector and tensor
polarizabilities for any spin-$S$ target and apply them to the
spin-1 deuteron. We calculate these polarizabilities of the
deuteron to the next-to-leading order in the pionless effective
field theory.
\end{abstract}

\maketitle

Low-energy photon scattering on a composite system can be
characterized by a host of electromagnetic polarizabilities, many
of which depend on the polarization (spin) state of the system.
Comparing experimental measurements of these polarizabilities with
theoretical predictions allows one to learn about the underlying
dynamics of the composite system. In this paper, we are interested
in the {\it spin-dependent} vector and tensor polarizabilities of
the deuteron. The spin-independent electric polarizability
$\alpha_{E0}$ of the nucleus has long been a subject of
investigation in the literature, and it has been explored
extensively in the potential models \cite{pot}. More recently, it
has been calculated in effective field theories with and without
explicit pion degrees of freedom \cite{chen1,chen2}.
Experimentally, $\alpha_{E0}$ has been measured through deuteron
scattering off heavy atoms ($0.70\pm 0.05$ fm$^3$) \cite{rodning}
and also extracted from the photo-production data through a sum
rule ($0.69\pm 0.04$ fm$^3$) \cite{friar}.

We communicate two sets of results in this paper. First is the sum
rules for the spin-dependent vector and tensor polarizabilities.
Because the deuteron binding-energy is 2.2 MeV, extracting the
polarizabilities directly from Compton scattering is difficult
experimentally. One would need high-intensity, polarized photon
beams at energy of order 1 MeV or less, which are not available at
the present time. [Compton scattering on the deuteron has been
studied in the past and sum rules have been explored
\cite{weyrauch}, and recently it has been investigated in
effective field theories \cite{compton}.] One could scatter a
polarized deuteron beam off the Coulomb field of a heavy nucleus,
as in \cite{rodning}, observing spin-dependent effects. However,
an easier way might be to extract these polarizabilities from the
spin-dependent photo-production cross sections through sum rules.
The HIGS facility at Duke could be used for this purpose
\cite{higs}.

The second set of results is on the effective field theory (EFT)
calculations of the forward spin polarizabilities. A systematic
EFT approach to the deuteron structure and scattering processes
has been developed in the last few years \cite{KSW} and has been
applied successfully to many experimental observables (see
\cite{review} for a review). Here we use the pionless version of
the theory \cite{chen2} to calculate the vector and tensor
polarizabilities up to the next-to-leading order. Since all the
counter terms to this order have been fixed from other processes,
there are no free parameters in our prediction.

Before specializing to the spin-1 deuteron case, we consider the
forward scattering of a circularly polarized photon of positive
helicity on a nucleus of spin $S$ and the magnetic quantum number
$m_S$ (we choose the direction of photon momentum as the
quantization axis $z$). The total number of forward scattering
amplitudes is easily found to be $2S+1+[S]$, where $[S]$ denotes
the integer part. These amplitudes arise from the initial deuteron
and photon states with the total angular momentum projection $S+1,
S, (S-1)^2,..., 1^2, 0^2$ for integer nuclear spin, and $S+1, S,
(S-1)^2,..., (3/2)^2, (1/2)^2$ for half integer nuclear spin,
where the superscripts denote the multiplicity of the amplitudes.
Here we are concerned with the forward amplitudes without helicity
flip, for which there are exactly $2S+1$.

Let us denote the scattering amplitude for $\gamma(+1)+ A(m_S)
\rightarrow \gamma(+1)+A(m_S)$ by $f^{(m_S)}(\theta)$, where
$\theta$ is the scattering angle, and the corresponding cross
section by $\sigma^{(m_S)}$. Then the well-known optical theorem
states
\begin{equation}
     {\rm Im} f^{(m_S)}(0) = \frac{k}{4\pi} \sigma^{(m_S)} \ .
\end{equation}
where $k$ is the center-of-mass momentum. Since $f^{(m_S)}\sim
\chi_{m_S}^*\chi_{m_S}$ where $\chi_{m_S}$ is the spin wave
function of the target, we can couple the initial and final spin
wave functions into tensors with definite total angular momentum,
\begin{equation}
      \tilde f_{J} = \frac{1}{\sqrt{2S+1}}\sum_{m_S} (-1)^{S-m_S}
            \langle S -m_S S m_S|J0\rangle f^{(m_S)} \ ,
\end{equation}
and a similar relation can be used to define $\tilde \sigma_{J}$
such that the optical theorem exists between them: ${\rm Im}
\tilde f_J(0) = \frac{k}{4\pi} \tilde \sigma_J$.
 Obviously for
$J=0$, one has
\begin{equation}
     \tilde f_{0} = \frac{1}{2S+1}\sum_{m_S}f^{(m_S)}\ ,
\end{equation}
which is just the unpolarized scattering amplitude.

On the other hand, a general scattering amplitude $f$ can be
expressed in terms of various non-relativistic structures after a
non-relativistic reduction. In forward scattering, it is
straightforward to write down all tensor structures for a general
spin target,
\begin{equation}
f = f_{0} \hat{\epsilon}^*\cdot \hat{\epsilon}
   + f_{1} i\hat{\epsilon}^* \times \hat{\epsilon} \cdot
   \vec{S}
   + f_{2}   (\hat k\otimes \hat k)^{(2)} \cdot (\vec S\otimes
   \vec S)^{(2)} \hat{\epsilon}^*\cdot \hat{\epsilon}
   + ...\
\end{equation}
where $\epsilon$ is the photon polarization, $\vec{S}$ is the
angular momentum operator of the target, and $\otimes$ indicates a
tensor coupling. [For example, $(\hat k\otimes \hat k)^{(2)} \cdot
(\vec S\otimes   \vec S)^{(2)} = (\hat k\cdot S)(\hat k\cdot
S)-S^2/3$.]  A general odd-$J$ term has the structure $i(\vec
S\otimes \vec S...\otimes \vec
S)^{(J)}\cdot((\hat\epsilon^*\times\hat\epsilon)\otimes\hat
k...\otimes\hat k)^{(J)}$; an even-$J$ term has the structure
$(\vec S\otimes \vec S...\otimes \vec S)^{(J)}\cdot (\hat k\otimes
\hat k... \otimes \hat k)^{(J)} \hat \epsilon^*\cdot\hat\epsilon$.
[If one considers the spin-flip forward amplitudes as well, one
has one more structure $(\vec{S}\otimes\vec{S}...\otimes
\vec{S})^{(J)}\cdot(\hat \epsilon^* \otimes \hat
\epsilon\otimes\hat k...\otimes \hat k)^{(J)}$ for every
even-$J$.] It is easy to show that the above amplitudes $f_i$ are
simply proportional to the ones in the previous section;
\begin{eqnarray}
   f_0 &=& \tilde f_0 \nonumber \\
   f_1 &=& -\sqrt{\frac{3}{S(S+1)}} \tilde f_1 \nonumber \\
   &=& -\frac{3}{S(S+1)}\frac{1}{2S+1}\sum_{m_S} m_Sf^{(m_S)}  \ ,
\end{eqnarray}
and
\begin{eqnarray}
   f_2 &=& \sqrt{\frac{5}{S(S+1)}}\frac{3}{\sqrt{(2S-1)(2S+3)}} \tilde f_2 \nonumber \\
   &=& \frac{3\cdot 5}{S(S+1)(2S-1)(2S+3)}\frac{1}{2S+1}\sum_{m_S}
   \left(3m_S^2-S(S+1)\right)f^{(m_S)}\ ,
\end{eqnarray}
and so on.

Because of the crossing symmetry under exchanging
$q^\mu\leftrightarrow -q^\mu$ and $\epsilon^{'*}\leftrightarrow
\epsilon$, all even-$J$ amplitudes are even functions of the
photon energy $\omega$, and all odd-$J$ amplitudes are odd
functions of $\omega$. Using the analyticity of the $f_{J}$, we
write down once-subtracted dispersion relations for even-$J$,
\begin{equation}
    f_{J}(\omega) = f_{J}(0) + \frac{2}{\pi}\omega^2
       \int^\infty_0 \frac{d\omega'}{\omega'}\frac{{\rm Im}
       f_{J}(\omega')}{{\omega'}^2-{\omega}^2}\ .
\end{equation}
Using the optical theorem, one has,
\begin{equation}
    f_{J}(\omega) = f_{J}(0) + \frac{\omega^2}{2\pi^2}
       \int^\infty_0
       {d\omega'}\frac{\sigma_{J}(\omega')}{{\omega'}^2-{\omega}^2}\
       ,
\end{equation}
which is the basis for various sum rules.

Consider the example of $J=0$, the low-energy expansion for the
amplitude is
\begin{equation}
    f_{0} = -\frac{e^2Z^2}{4\pi M} + (\alpha_{E0} +
    \beta_{M0})\omega^2 + ... \ .
\end{equation}
Substituting the above into the dispersion relation, we recover
the well-known Baldin sum rule for the averaged cross section,
\begin{equation}
  \alpha_{E0} + \beta_{M0} =  \frac{1}{2\pi^2}
       \int^\infty_0
       {d\omega'}\frac{\sigma_{0}(\omega')}{{\omega'}^2} \
\end{equation}
For the special case of the spin-1 deuteron, the sum rule becomes
\begin{equation}
     \alpha_{E0} + \beta_{M0} =  \frac{1}{6\pi^2}  \int^\infty_0
       {d\omega'}\frac{\sigma^{(1)}+\sigma^{(0)} +
       \sigma^{(-1)}}{{\omega'}^2}\ ,
\end{equation}
where we remind the reader that $\sigma^{(m)}$ denotes the cross
section for the deuteron in an $m$-state.

Let us review the calculations of $\alpha_{E0}$ and $\beta_{M0}$
in the pionless effective field theory \cite{chen1,
chen2,phillips}. We remind the reader that in the pionless theory,
the leading order effective lagrangian is
\begin{eqnarray}
  {\cal L} &=& N^\dagger \left(iD_0 +
  \frac{\vec{D}^2}{2M_N}\right)N
    - C_0^{(^3S_1)}(\mu) (N^T P_iN)^\dagger (N^TP_i N) \nonumber \\
    && - C_0^{(^1S_0)}(\mu) (N^T \overline{P}_iN)^\dagger (N^T\overline{P}_i N)
    + \frac{e}{2M_N}N^\dagger(\mu^{(0)}+
  \mu^{(1)}\tau_3 )\vec{\sigma}\cdot \vec{B} N \ ,
\end{eqnarray}
where $N$ is the nucleon field,
$P_i=\tau_2\sigma_2\sigma_i/\sqrt{8}$ and
$\overline{P}_i=\sigma_2\tau_2\tau_i/\sqrt{8}$ are the triplet
$S_1$ and singlet $S_0$ two-nucleon projection operators,
respectively. The covariant derivative is $\vec{D}= \vec{\partial}
+ ieQ\vec{A}$ with $Q=(1+\tau^3)/2$ as the charge operator and
$\vec{A}$ the photon vector potential. $\mu^{(0)}=(\mu_p+\mu_n)/2$
and $\mu^{(1)}=(\mu_p-\mu_n)/2$ are the isoscalar and isovector
nucleon magnetic moments in nuclear magnetons. The two-body
coupling constants are
\begin{eqnarray}
   C_0^{(^1S_0)}(\mu) &=& -\frac{4\pi}{M_N}
          \frac{1}{\left(\mu-1/a^{(^1S_0)}\right)}\   \nonumber \\
    C_0^{(^3S_1)}(\mu) &=&-
    \frac{4\pi}{M_N}\frac{1}{(\mu-\gamma)} \  ,
\end{eqnarray}
where $\mu$ is a renormalization scale, $a^{(^1S_0)}=-23.714$ fm
is the scattering length in the two-nucleon singlet $S_0$ channel
and $\gamma=\sqrt{M_NB}=45.703$ MeV with $B=2.225$ MeV the
deuteron binding energy.

To the N$^3$LO order, the scalar electric polarizability is
\cite{chen2,phillips}
\begin{equation}
     \alpha_{E0} = \frac{\alpha_{\rm em} M_N}{32\gamma^4} Z_d
     \left[1+\frac{2 \gamma^2}{3M_N^2} + \frac{M_N\gamma^3}{3\pi}
     D_P\right] \ ,
\end{equation}
where $Z_d =1/(1-\gamma\rho_d)= 1.69$ is the deuteron wave
function renormalization; $\rho_d=1.764$ fm; $D_p=-1.51$ fm$^3$.
Numerically $\alpha_{E0} = 0.6339$ fm$^3$. The magnetic
polarizability $\beta_{M0}$ is suppressed by two orders of
$Q$-counting relative to $\alpha_{E0}$ because of the
non-relativistic dynamics of the deuteron \cite{chen1},
\begin{equation}
    \beta_{M0} = -\frac{\alpha_{\rm em}}{32\gamma^2M_N}
    \left[1-\frac{16(\mu^{(1)})^2}{3} - \frac{8\gamma M_N
    (\mu^{(1)})^2}{3\pi}{\cal A}^{(^1S_0)}_{-1}(-B)\right]\ ,
\end{equation}
where ${\cal A}_{-1}^{(^1S_0)}(-B)= -
(4\pi/M_N)(1/a^{(^1S_0)}-\gamma)^{-1}$ is the leading-order
singlet $S_0$ scattering amplitude at energy $E=-B$. [Note that
the above result differs from \cite{chen1} by a term proportional
to $(\mu^{(0)})^2$, which is cancelled by an omitted diagram with
the triplet $S_1$ bubble chain between the two photon insertions.]
Numerically, $\beta_{M0} = 0.067$ fm$^3$, about 10\% of
$\alpha_{E0}$.

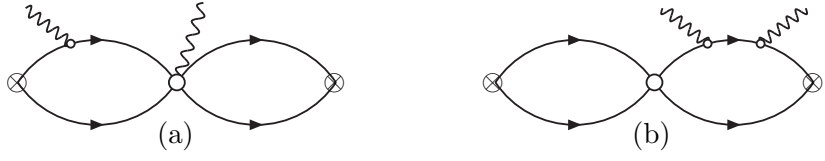
\begin{figure}
\SetWidth{0.7}
\begin{center}

\begin{picture}(420,80)(0,0)

\Text(60,15)[]{$\otimes$} \ArrowArcn(90,-5)(36,146.4,33.6)
\ArrowArc(90,35)(36,213.6,326.4) \ArrowArcn(150,-5)(36,146.4,33.6)
\ArrowArc(150,35)(36,213.6,326.4) \Photon(63,43)(80,29.6){2}{5}
\Photon(130,45)(120,15){-2}{5} \BCirc(80,29.6){1.5}
\BCirc(120,15){3} \Text(180,15)[]{$\otimes$}

\Text(240,15)[]{$\otimes$} \ArrowArcn(270,-5)(36,146.4,33.6)
\ArrowArc(270,35)(36,213.6,326.4)
\ArrowArcn(330,-5)(36,146.4,33.6)
\ArrowArc(330,35)(36,213.6,326.4) \Photon(357,43)(340,29.6){2}{5}
\Photon(303,43)(320,29.6){-2}{5} \BCirc(340,29.6){1.5}
\BCirc(320,29.6){1.5} \BCirc(300,15){3} \Text(361,15)[]{$\otimes$}

\Text(120,-5)[]{(a)} \Text(300,-5)[]{(b)}

\end{picture}

\end{center}

\caption{Leading-order contribution to the electric tensor
polarizability of the deuteron. The small circle denotes the
electric current coupling. The large circle represents the S-D
mixing interaction $C_0^{(sd)}$. The crossing circles represent
the deuteron interpolating fields.}
\end{figure}

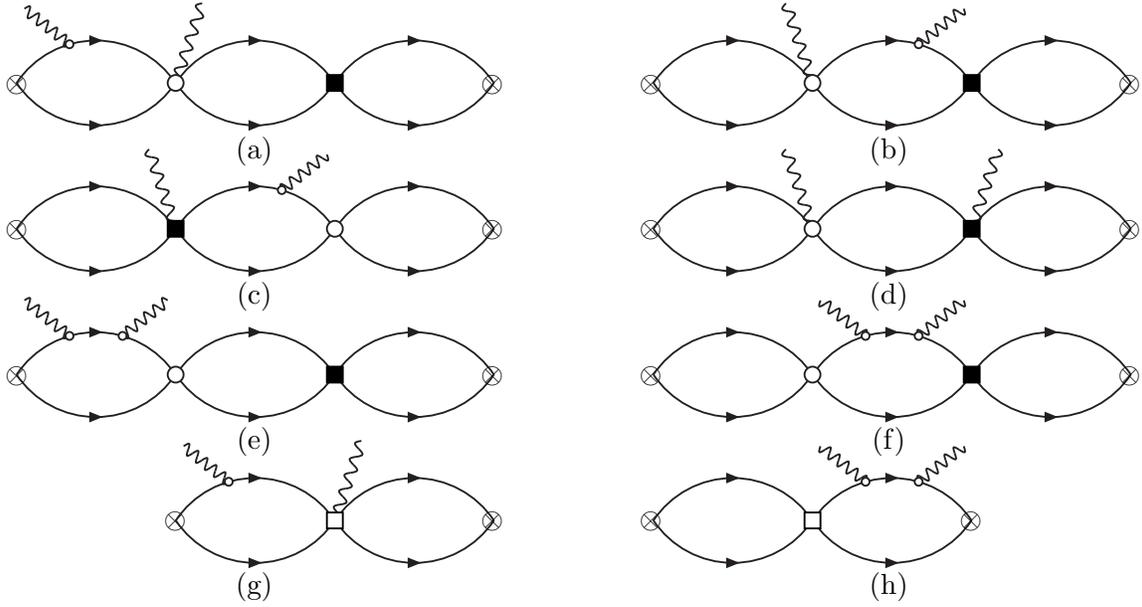
\begin{figure}
\SetWidth{0.7}
\begin{center}

\begin{picture}(420,220)(0,0)

\Text(0,180)[]{$\otimes$} \ArrowArcn(30,160)(36,146.4,33.6)
\ArrowArc(30,200)(36,213.6,326.4)
\ArrowArcn(90,160)(36,146.4,33.6)
\ArrowArc(90,200)(36,213.6,326.4)
\ArrowArcn(150,160)(36,146.4,33.6)
\ArrowArc(150,200)(36,213.6,326.4) \Photon(3,208)(20,194.6){2}{5}
\Photon(70,210)(60,180){-2}{5} \BCirc(20,194.6){1.5}
\BCirc(60,180){3} \GBoxc(120,180)(6,6){0}
\Text(180,180)[]{$\otimes$}

\Text(240,180)[]{$\otimes$} \ArrowArcn(270,160)(36,146.4,33.6)
\ArrowArc(270,200)(36,213.6,326.4)
\ArrowArcn(330,160)(36,146.4,33.6)
\ArrowArc(330,200)(36,213.6,326.4)
\ArrowArcn(390,160)(36,146.4,33.6)
\ArrowArc(390,200)(36,213.6,326.4)
\Photon(357,208)(340,194.6){2}{5} \Photon(290,210)(300,180){-2}{5}
\BCirc(340,194.6){1.5} \BCirc(300,180){3} \GBoxc(360,180)(6,6){0}
\Text(421,180)[]{$\otimes$}

\Text(0,125)[]{$\otimes$} \ArrowArcn(30,105)(36,146.4,33.6)
\ArrowArc(30,145)(36,213.6,326.4)
\ArrowArcn(90,105)(36,146.4,33.6)
\ArrowArc(90,145)(36,213.6,326.4)
\ArrowArcn(150,105)(36,146.4,33.6)
\ArrowArc(150,145)(36,213.6,326.4)
\Photon(117,153)(100,139.6){2}{5} \Photon(50,155)(60,125){-2}{5}
\BCirc(100,139.6){1.5} \BCirc(120,125){3} \GBoxc(60,125)(6,6){0}
\Text(180,125)[]{$\otimes$}

\Text(240,125)[]{$\otimes$} \ArrowArcn(270,105)(36,146.4,33.6)
\ArrowArc(270,145)(36,213.6,326.4)
\ArrowArcn(330,105)(36,146.4,33.6)
\ArrowArc(330,145)(36,213.6,326.4)
\ArrowArcn(390,105)(36,146.4,33.6)
\ArrowArc(390,145)(36,213.6,326.4) \Photon(370,155)(360,125){2}{5}
\Photon(290,155)(300,125){-2}{5} \BCirc(300,125){3}
\GBoxc(360,125)(6,6){0} \Text(421,125)[]{$\otimes$}

\Text(0,70)[]{$\otimes$} \ArrowArcn(30,50)(36,146.4,33.6)
\ArrowArc(30,90)(36,213.6,326.4) \ArrowArcn(90,50)(36,146.4,33.6)
\ArrowArc(90,90)(36,213.6,326.4) \ArrowArcn(150,50)(36,146.4,33.6)
\ArrowArc(150,90)(36,213.6,326.4) \Photon(3,98)(20,84.6){2}{5}
\Photon(57,98)(40,84.6){-2}{5} \BCirc(20,84.6){1.5}
\BCirc(40,84.6){1.5} \BCirc(60,70){3} \GBoxc(120,70)(6,6){0}
\Text(180,70)[]{$\otimes$}

\Text(240,70)[]{$\otimes$} \ArrowArcn(270,50)(36,146.4,33.6)
\ArrowArc(270,90)(36,213.6,326.4)
\ArrowArcn(330,50)(36,146.4,33.6)
\ArrowArc(330,90)(36,213.6,326.4)
\ArrowArcn(390,50)(36,146.4,33.6)
\ArrowArc(390,90)(36,213.6,326.4) \Photon(357,98)(340,84.6){2}{5}
\Photon(303,98)(320,84.6){-2}{5} \BCirc(340,84.6){1.5}
\BCirc(320,84.6){1.5} \BCirc(300,70){3} \GBoxc(360,70)(6,6){0}
\Text(421,70)[]{$\otimes$}

\Text(60,15)[]{$\otimes$} \ArrowArcn(90,-5)(36,146.4,33.6)
\ArrowArc(90,35)(36,213.6,326.4) \ArrowArcn(150,-5)(36,146.4,33.6)
\ArrowArc(150,35)(36,213.6,326.4) \Photon(63,43)(80,29.6){2}{5}
\Photon(130,45)(120,15){-2}{5} \BCirc(80,29.6){1.5}
\GBoxc(120,15)(6,6){1} \Text(180,15)[]{$\otimes$}

\Text(240,15)[]{$\otimes$} \ArrowArcn(270,-5)(36,146.4,33.6)
\ArrowArc(270,35)(36,213.6,326.4)
\ArrowArcn(330,-5)(36,146.4,33.6)
\ArrowArc(330,35)(36,213.6,326.4) \Photon(357,43)(340,29.6){2}{5}
\Photon(303,43)(320,29.6){-2}{5} \BCirc(340,29.6){1.5}
\GBoxc(300,15)(6,6){1} \BCirc(320,29.6){1.5}
\Text(361,15)[]{$\otimes$}

\Text(90,155)[]{(a)} \Text(330,155)[]{(b)} \Text(90,100)[]{(c)}
\Text(330,100)[]{(d)} \Text(90,45)[]{(e)} \Text(330,45)[]{(f)}

\Text(90,-10)[]{(g)} \Text(330,-10)[]{(h)}

\end{picture}

\end{center}

\caption{The next-to-leading order contribution to the electric
tensor polarizability of the deuteron. The black box represents
the triplet $C_2^{(^3S_1)}$. The small circle denotes the electric
current coupling. The large circle represents the S-D mixing
interaction $C_0^{(sd)}$. The blank square box represents the NLO
S-D mixing vertex: $C_2^{(sd)}$ and $\widetilde{C}_2^{(sd)}$. The
crossing diagrams are not shown. }
\end{figure}

Turning to the case of $J=2$, we have the low-energy expansion,
\begin{equation}
   f_{2}(\omega) = (\alpha_{E2} + \beta_{M2})\omega^2 + {\cal
   O}(\omega^4) \ ,
\end{equation}
which allows us to write a $J=2$ sum rule,
\begin{equation}
  \alpha_{E2} + \beta_{M2} =
  \frac{1}{2\pi^2}\int^\infty_0 d\omega'
  \frac{\sigma_2}{{\omega'}^2} \ .
\end{equation}
Specializing to the spin-1 deuteron, the sum rule becomes,
\begin{equation}
  \alpha_{E2} + \beta_{M2} = \frac{1}{4\pi^2}
       \int^\infty_0
       {d\omega'}\frac{\sigma^{(1)}+\sigma^{(-1)}-2\sigma^{(0)}}{{\omega'}^2}
       \ .
\end{equation}
In the effective theory with pions, the leading contribution comes
at the NLO in Q-counting from the potential pion exchange
\cite{chen1,compton}. On the other hand, in the pionless theory,
there is a leading contribution coming from the two-body operator
that couples the single $S_0$ and triplet $S_1$ channels
\cite{chen3},
\begin{equation}
   {\cal L} = - {\cal T}^{(sd)}_{ij,xy} C_0^{sd} (N^TP^iN)^\dagger
   \left(N^T{\cal O}_2^{xy,j}N\right) \ ,
\end{equation}
where ${\cal T}^{(sd)}_{ij, xy} = \delta_{ix}\delta_{jy}  -
\delta_{ij}\delta_{xy}/(n-1)$ and ${\cal O}_2^{xy, j} =
-(D^xD^yP^j + P^j D^x D^y - D^xP^jD^y-D^yP^j D^x)/4$. $C_0^{sd}$
is related to the asymptotic D/S ratio $\eta_{sd}$ of the deuteron
(0.0254) through \cite{chen2}
\begin{equation}
          C_0^{(sd)} = -\eta_{sd}
          \frac{6\sqrt{2}\pi}{M_N\gamma^2(\mu-\gamma)} \ .
\end{equation}
The leading-order Feynman diagrams are shown in Fig. 1, and a
straightforward calculation yields,
\begin{equation}
        \alpha_{E2}^{\rm LO} = - \frac{3\sqrt{2}\alpha_{\rm em}\eta_{sd}
        M_N}{32\gamma^4} \ .
\end{equation}

At the NLO, there are contributions from Fig. 2 plus the
correction for wave function renormalization. The result is that
\begin{eqnarray}
        \alpha_{E2}^{\rm LO+NLO} &=& - \frac{3\sqrt{2}\alpha_{\rm em}\eta_{sd}
        M_N}{32\gamma^4} ( 1+ \gamma\rho_d) \nonumber \\
            &\approx &  - \frac{3\sqrt{2}\alpha_{\rm em}\eta_{sd}
        M_N}{32\gamma^4} Z_d \ ,
\end{eqnarray}
where in the second line we have introduced $Z_d
=1/(1-\gamma\rho_d)= 1.69$ \cite{phillips}. Numerically, we have
$\alpha_{E2} = -0.068 $ fm$^3$ at this order, which is very close
to the result from the potential pion contribution \cite{chen1}.
For completeness, we quote the magnetic tensor polarizability
$\beta_{M2}$ which formally comes at N$^2$LO \cite{chen1},
\begin{equation}
   \beta_{M2} = \frac{\alpha_{\rm
   em}(\mu^{(1)})^2}{2M_N\gamma^2}\left[1 +
   \frac{M_N\gamma}{2\pi} {\cal A}^{(^1S_0)}_{-1}(B)\right] \ .
\end{equation}
However, it is very large numerically, 0.195 fm$^3$, because of
the large isovector magnetic moment. [Again, the $(\mu^{(0)})^2$
dependence in Ref. \cite{chen1} should be absent.]

 For odd-$J$, one can write down a dispersion relation without
subtraction,
\begin{equation}
    f_{J}(\omega) = \frac{2}{\pi}\omega
       \int^\infty_0 d\omega' \frac{{\rm Im}
       f_{J}(\omega')}{{\omega'}^2-{\omega}^2}\ .
\end{equation}
Again using the optical theorem, one has
\begin{equation}
    f_{J}(\omega) = \frac{\omega}{2\pi^2}
       \int^\infty_0 d\omega'\omega'
       \frac{\sigma_{J}(\omega')}{{\omega'}^2-{\omega}^2}\ .
\label{odis}
\end{equation}
For $J=1$, the scattering amplitude has a low-energy expansion,
\begin{equation}
f_{1} = - \frac{\alpha_{\rm em}\kappa^2}{4S^2M^2}\omega + 2\gamma
\omega^3 + ...
\end{equation}
where the first term corresponds to the famous low-energy theorem
\cite{low} with the anomalous magnetic moment $\kappa$ defined as
$\mu-2S$ \cite{low1}, where $\mu$ is the magnetic moment in unit
of $e\hbar/2Mc$. The next term defines the forward
spin-polarizability $\gamma$. Substituting the above into Eq.
(\ref{odis}), the first term yields the famous Drell-Hearn
Gerasimov (DHG) sum rule, now extended to target of any spin $S$,
\begin{equation}
    \frac{\alpha_{\rm em}\kappa^2}{4S^2M^2} = \frac{1}{2\pi^2}
       \int^\infty_0 d\omega' \frac{\sigma_{1}(\omega')}{\omega'}
       \ ,
\end{equation}
where $\sigma_1 = [3/S(S+1)(2S+1)]\sum_{m_s} m_S \sigma_{m_S}$.
For a discussion about the DHG sum rule for the deuteron, see Ref.
\cite{arenhovel}.

The sum rule for the forward spin polarizability reads
\begin{equation}
   2\gamma = \frac{1}{2\pi^2}
       \int^\infty_0 d\omega' \frac{\sigma_1(\omega')}{\omega'^3}
       \  .
\end{equation}
Specializing to the deuteron, we have
\begin{equation}
   \gamma = -\frac{1}{8\pi^2}
       \int^\infty_0 d\omega'
       \frac{\sigma^{(1)}-\sigma^{(-1)}}{\omega'^3}\ .
\end{equation}
This is the sum rule potentially useful for extracting the forward
spin polarizability from the polarized photo-production cross
section.

\begin{figure}
\SetWidth{0.7}
\begin{center}

\begin{picture}(420,80)(0,0)
\ArrowArcn(40,-10)(50,143,37) \ArrowArc(40,50)(50,217,323)
\Text(0,20)[]{$\otimes$} \Text(80,20)[]{$\otimes$}
\Photon(10,52.7)(25,37.7){-2}{5} \Photon(70,52.7)(55,37.7){-2}{5}
\GCirc(25,37.7){2}{.7} \GCirc(55,37.7){2}{.7}

\Text(120,20)[]{$\otimes$} \ArrowArcn(150,0)(36,146.4,33.6)
\ArrowArc(150,40)(36,213.6,326.4) \ArrowArcn(210,0)(36,146.4,33.6)
\ArrowArc(210,40)(36,213.6,326.4) \Vertex(252,20){1}
\Vertex(264,20){1} \Vertex(276,20){1} \Vertex(288,20){1}
\ArrowArcn(330,0)(36,146.4,33.6) \ArrowArc(330,40)(36,213.6,326.4)
\ArrowArcn(390,0)(36,146.4,33.6) \ArrowArc(390,40)(36,213.6,326.4)
\Text(421,20)[]{$\otimes$} \Photon(123,49)(138,34){-2}{5}
\Photon(417,49)(402,34){-2}{5} \GCirc(138,34){2}{.7}
\GCirc(402,34){2}{.7} \Text(40,-15)[]{(a)} \Text(270,-15)[]{(b)}
\end{picture}
\end{center}
\caption{Leading order contribution to the forward vector
polarizability of deuteron. The gray circle denotes the magnetic
moment coupling. The crossing diagrams are not shown. The chain
bubble in (b) includes insertions of both triplet and singlet
types of $C_0$: $C_0^{(^3S_1)}$ and $C_0^{(^1S_0)}$. }
\end{figure}
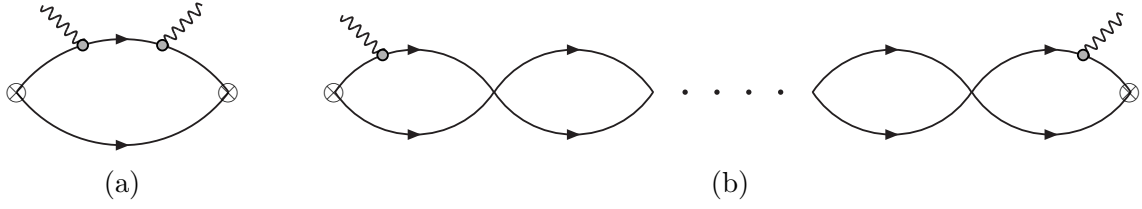

The relevant Feynman diagrams for the forward Compton scattering
are shown in Fig. 3. Taking into account the crossing symmetry,
the result for the leading-order spin polarizability is
\begin{eqnarray}
     \gamma^{\rm LO} = \frac{\alpha_{\rm em}
     (\mu^{(1)})^2}{16\gamma^4}\left[1 + \frac{M_N\gamma}{2\pi}{\cal A}_{-1}^{(^1S_0)}(-B)
     \left(1 + \frac{M_N\gamma}{4\pi}{\cal A}_{-1}^{(^1S_0)}(-B)\right)
     \right]\ .
\end{eqnarray}
The numerical value of $\gamma$ at this order is $3.596$ fm$^4$.

\begin{figure}
\SetWidth{0.7}
\begin{center}

\begin{picture}(400,250)(0,0)
\ArrowArcn(40,180.3)(60,131.8,48.2)
\ArrowArc(40,269.7)(60,228.2,311.8) \Text(0,226)[]{$\otimes$}
\GBoxc(80,225)(6,6){0} \Photon(10,253.4)(25,238.4){-2}{5}
\Photon(70,253.4)(55,238.4){-2}{5}
\ArrowArcn(120,180.3)(60,131.8,48.2)
\ArrowArc(120,269.7)(60,228.2,311.8) \Text(161,226)[]{$\otimes$}
\GCirc(25,238.4){2}{.7} \GCirc(55,238.4){2}{.7}

\ArrowArcn(280,180.3)(60,131.8,48.2)
\ArrowArc(280,269.7)(60,228.2,311.8) \Text(241,226)[]{$\otimes$}
\GBoxc(320,225)(6,6){0} \Photon(250,253.4)(265,238.4){-2}{5}
\Photon(390,253.4)(375,238.4){-2}{5}
\ArrowArcn(360,180.3)(60,131.8,48.2)
\ArrowArc(360,269.7)(60,228.2,311.8) \Text(401,226)[]{$\otimes$}
\GCirc(265,238.4){2}{.7} \GCirc(375,238.4){2}{.7}

\ArrowArcn(40,130.3)(60,131.8,48.2)
\ArrowArc(40,219.7)(60,228.2,311.8) \Text(0,176)[]{$\otimes$}
\GBoxc(80,175)(6,6){0} \Photon(10,203.4)(25,188.4){-2}{5}
\ArrowArcn(120,130.3)(60,131.8,48.2)
\ArrowArc(120,219.7)(60,228.2,311.8) \Vertex(176,175){1}
\Vertex(192,175){1} \Vertex(208,175){1} \Vertex(224,175){1}

\ArrowArcn(280,130.3)(60,131.8,48.2)
\ArrowArc(280,219.7)(60,228.2,311.8)
\Photon(390,203.4)(375,188.4){-2}{5}
\ArrowArcn(360,130.3)(60,131.8,48.2)
\ArrowArc(360,219.7)(60,228.2,311.8) \Text(401,176)[]{$\otimes$}
\GCirc(25,188.4){2}{.7} \GCirc(375,188.4){2}{.7}

\ArrowArcn(40,80.3)(60,131.8,48.2)
\ArrowArc(40,169.7)(60,228.2,311.8) \Text(0,126)[]{$\otimes$}
\GBoxc(80,125)(6,6){0} \Photon(90,153.4)(105,138.4){-2}{5}
\ArrowArcn(120,80.3)(60,131.8,48.2)
\ArrowArc(120,169.7)(60,228.2,311.8) \Vertex(176,125){1}
\Vertex(192,125){1} \Vertex(208,125){1} \Vertex(224,125){1}

\ArrowArcn(280,80.3)(60,131.8,48.2)
\ArrowArc(280,169.7)(60,228.2,311.8)
\Photon(390,153.4)(375,138.4){-2}{5}
\ArrowArcn(360,80.3)(60,131.8,48.2)
\ArrowArc(360,169.7)(60,228.2,311.8) \Text(401,126)[]{$\otimes$}
\GCirc(105,138.4){2}{.7} \GCirc(375,138.4){2}{.7}

\ArrowArcn(40,30.3)(60,131.8,48.2)
\ArrowArc(40,119.7)(60,228.2,311.8) \Text(0,76)[]{$\otimes$}
\GBoxc(200,75)(6,6){0} \Photon(10,103.4)(25,88.4){-2}{5}
\ArrowArcn(160,30.3)(60,131.8,48.2)
\ArrowArc(160,119.7)(60,228.2,311.8)

\Vertex(90,75){1} \Vertex(100,75){1} \Vertex(110,75){1}
\Vertex(290,75){1} \Vertex(300,75){1} \Vertex(310,75){1}

\ArrowArcn(240,30.3)(60,131.8,48.2)
\ArrowArc(240,119.7)(60,228.2,311.8)
\Photon(390,103.4)(375,88.4){-2}{5}
\ArrowArcn(360,30.3)(60,131.8,48.2)
\ArrowArc(360,119.7)(60,228.2,311.8) \Text(401,76)[]{$\otimes$}
\GCirc(25,88.4){2}{.7} \GCirc(375,88.4){2}{.7}
\ArrowArcn(40,-20.3)(60,131.8,48.2)
\ArrowArc(40,69.7)(60,228.2,311.8) \Text(1,25)[]{$\otimes$}
\GCirc(80,25){3}{0} \Photon(10,53.4)(25,38.4){-2}{5}
\Photon(95,45)(80,25){-2}{5} \ArrowArcn(120,-20.3)(60,131.8,48.2)
\ArrowArc(120,69.7)(60,228.2,311.8) \Text(160,25)[]{$\otimes$}
\GCirc(25,38.4){2}{.7} \Text(80,205)[]{(a)}

\Text(320,205)[]{(b)}

\Text(200,155)[]{(c)}

\Text(200,105)[]{(d)}

\Text(200,55)[]{(e)}

\Text(80,5)[]{(f)}

\end{picture}

\end{center}

\caption{The next-to-leading order contribution to the magnetic
vector polarizability of deuteron. The square black box represent
both triplet and singlet types of $C_2$: $C_2^{(^3S_1)}$ and
$C_2^{(^1S_0)}$. The large black dot denotes the $L_1$ coupling.
The gray circle denotes the magnetic moment coupling. The crossing
diagrams are not shown. The chain bubble in (b) includes
insertions of both triplet and singlet types of $C_0$:
$C_0^{(^3S_1)}$ and $C_0^{(^1S_0)}$. }
\end{figure}
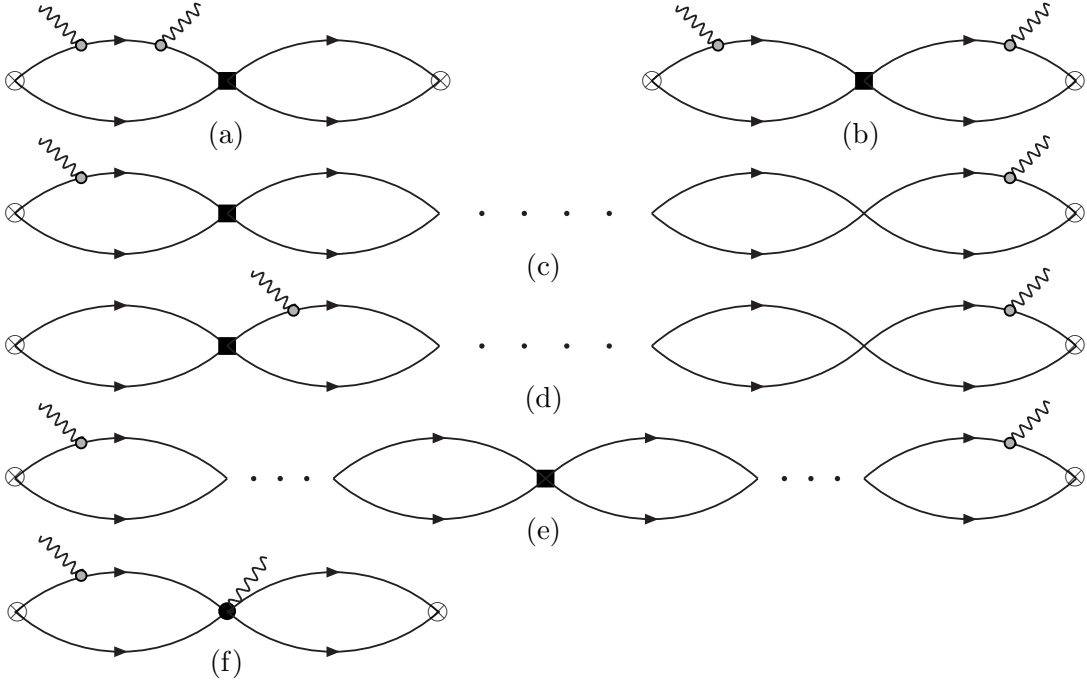

At the next-to-leading order, there are contributions from the
$C_2$ coupling in the singlet and triplet channels,
\begin{eqnarray}
    {\cal L}' &=&
    -C_2^{(^1S_0)}(\mu)\frac{1}{8}\left[(N^T\overline{P}_iN)^\dagger
    \left(N^T \left[\overline{P}_i \overrightarrow{D}^2 +\overleftarrow{D}^2\overline{P}_i
    - 2\overleftarrow{D}\overline{P}_i\overrightarrow{D}\right]N\right)
     + {\rm h. c.}\right] \nonumber \\
    && -C_2^{(^3S_1)}(\mu)\frac{1}{8}\left[(N^TP_iN)^\dagger
    \left(N^T \left[P_i \overrightarrow{D}^2 +\overleftarrow{D}^2P_i - 2\overleftarrow{D}
    P_i\overrightarrow{D}\right]N\right) + {\rm h. c.}\right] \ .
\end{eqnarray}
where
\begin{eqnarray}
    C_2^{(^1S_0)}(\mu) &=& \frac{4\pi}{M_N} \frac{r_0}{2}
          \frac{1}{\left(\mu-1/a^{(^1S_0)}\right)^2}\   \nonumber \\
    C_2^{(^3S_1)}(\mu) &=&
    \frac{2\pi}{M_N}\frac{\rho_d}{(\mu-\gamma)^2} \ ,
\end{eqnarray}
with $r_0=2.73$ fm. There are also contributions from the
following electromagnetic counter term \cite{wise}
\begin{equation}
    {\cal L}'' = eL_1 (N^TP_iN)^\dagger (N^T \overline{P}_3N)B_i \
    .
\end{equation}
The relevant next-to-leading order Feynman diagrams for the
Compton amplitude are shown in Fig. 2. The calculated vector
polarizability is
\begin{eqnarray}
     \gamma^{\rm NLO} &=& \frac{\alpha_{\rm em}
     (\mu^{(1)})^2M_N^2C_2^{(^1S_0)}(\mu)}{2(8\pi\gamma)^2}{\cal A}_{-1}^{(^1S_0)}(-B)
     \left(\mu-1/a^{(^1S_0)}\right)^2 \nonumber \\ &&\times \left[\frac{3\gamma-\mu}{\mu-1/a^{(^1S_0)}}
     - \frac{3M_N\gamma}{4\pi}{\cal A}_{-1}^{(^1S_0)}(-B)
      + \frac{(M_N\gamma)^2}{8\pi^2} {\cal A}_{-1}^{(^1S_0)}(-B)^2 \right]
     \nonumber \\
  &&  + \frac{\alpha_{\rm em}
     (\mu^{(1)})^2M_NC_2^{(^3S_1)}(\mu)}{32\pi\gamma^3}(\mu-\gamma)
     \left[\mu-2\gamma + (\mu-3\gamma)\frac{M_N\gamma}{4\pi}{\cal A}_{-1}^{(^1S_0)}(-B)
      \right] \nonumber \\
 && + \frac{\alpha_{\rm em}
     \mu^{(1)}L_1(\mu)M_N^2}{(8\pi)^2\gamma^2}(\mu-\gamma)\left(\mu-1/a^{(^1S_0)}\right)
    {\cal A}_{-1}^{(^1S_0)}(-B)
     \left[1 + \frac{M_N\gamma}{2\pi}{\cal
     A}_{-1}^{(^1S_0)}(-B)\right]
\end{eqnarray}
The last term in the first square bracket can be added to the
leading-order result if one replaces ${\cal A}_{-1}^{(^1S_0)}(-B)$
with $ {\cal A}_{-1}^{(^1S_0)}(-B) + {\cal A}_{0}^{(^1S_0)}(-B)$.
It is easy to verify that if the following equation holds,
\begin{equation}
     \mu\frac{d}{d\mu} \left[\frac{L_1(\mu)
     - \frac{1}{2}\mu^{(1)}(C_2^{(^1S_0)}(\mu)+C_2^{(^3S_1)}
     (\mu))}{C_0^{(^1S_0)}(\mu)C_0^{(^3S_1)}(\mu)}\right]=0 \ ,
\end{equation}
the above result is independent of the renormalization scale
$\mu$. This is consistent with the $np$ capture calculation
\cite{wise}.

Using the counter term determined from the neutron capture, $L_1 =
3.26$ fm$^4$, we find $\gamma^{\rm NLO}=0.324$ fm$^4$ which is
10\% of the leading-order result. Therefore the effective field
theory expansion seems to converge well.

In summary, we have presented sum rules for the vector and tensor
polarizabilities of the deuteron. We have evaluated both
quantities in the pionless effective field theory to the
next-to-leading order. The $Q$-expansion seems to converge well
for the vector polarizability.

We thank Jiunn-Wei Chen and Barry Holstein for discussions related
to the subject of this paper. This work was supported by the U. S.
Department of Energy via grant DE-FG02-93ER-40762.

\end{document}